\documentclass[12pt]{article}
\usepackage{amsfonts}
\usepackage{graphicx}
\usepackage{epstopdf}
\usepackage{epsfig}
\usepackage{amssymb}
\usepackage{setspace}
\usepackage{caption}
\usepackage{amsfonts}
\usepackage{color}
\usepackage{amsmath}
\usepackage{float}
\usepackage{comment}
\newcommand {\be}{\begin{equation}}
\newcommand {\ee}{\end{equation}}
 \newcommand {\bea}{\begin{array}}
 
 \newcommand {\eea}{\end{array}}

 \newcommand {\RN}{Reissner-Nordstrom~}

\textwidth=6.5in \hoffset=-.75in \voffset=-1in \numberwithin{equation}{section}
\numberwithin{figure}{section}

\begin{document}

\begin{titlepage}
	\vspace{1cm}
	\begin{center}
		{\Large \bf {Destroying Kaluza-Klein and Kerr-Newman black holes}}\\
	\end{center}
	\vspace{2cm}
	\begin{center}
		\renewcommand{\thefootnote}{\fnsymbol{footnote}}
		Haryanto M. Siahaan{\footnote{haryanto.siahaan@unpar.ac.id}} and Paulus C. Tjiang{\footnote{pctjiang@unpar.ac.id}}\\
		Center for Theoretical Physics,\\
		Department of Physics, Parahyangan Catholic University,\\
		Jalan Ciumbuleuit 94, Bandung 40141, Indonesia
		\renewcommand{\thefootnote}{\arabic{footnote}}
	\end{center}
	
	\begin{abstract}
		
We investigate the destroying of charged and rotating black holes in Einstein-Maxwell-(dilaton) theory. We show that a test particle with some appropriate properties in the black hole background can turn the black hole into a naked singularity. In this work, we neglect the self-force, self-energy, and radiative effects as considered by some others in literature. As the result, we are able to show that the Kerr-Newman and Kaluza-Klein black holes can be destroyed by the test particle. Furthermore, for Kaluza-Klein black hole, we consider the overspinning of the black hole by a neutral test scalar field.
		
	\end{abstract}
\end{titlepage}\onecolumn
\bigskip

\section{Introduction}
\label{sec:intro}

Black hole is a singularity covered by an event horizon with an extremely strong gravity around its surrounding. The advent of black hole image reported by the Event Horizon Telescope collaboration \cite{EventHorizonTelescope:2019dse,EventHorizonTelescope:2021dqv} confirms the prediction of black holes in Einstein gravity. This finding is a solid evidence of black hole existence which then invites further related studies on black hole \cite{Bozzola:2021elc,Narzilloev:2020peq,Alatas:2021oze}. Nevertheless, as the horizon ceases to exist, the singularity is then visible to an outside observer as a naked singularity. Penrose in 1969 proposed that no naked singularity can exist in our universe \cite{Penrose:1969pc}, and the proposal is known as the weak cosmic censorship conjecture (WCCC). This conjecture had been tested repeatedly, including the quite famous one by Wald in his 1974 gedanken experiment \cite{Wald}. In this Wald experiment of thought, one consider a black hole capturing a test particle which opens a possibility to turn into a naked singularity. However, in Wald's original scenario, it was shown that naked such naked singularity production is impossible.

About two decades after the Wald's gedanken experiment proposal \cite{Wald}, the conjecture by Penrose was again tested in the work by Hubeny \cite{Hubeny:1998ga} where she showed that a particular set up of a charged test object may turn a \RN black hole into a naked singularity. The similar idea to Hubeny's in violating WCCC for Kerr black hole was proposed Jacobson and Sotiriou \cite{Jacobson:2009kt}, for Kerr-Newman black hole were discussed in \cite{SaaPRD,Gao:2012ca}, and for magnetized Kerr black hole is investigated in \cite{Siahaan:2016zjw}. The reconsideration of WCCC in the fashion of Hubeny \cite{Hubeny:1998ga} is not limited just to black holes in Einstein-Maxwell theory, but also beyond \cite{Duztas:2020xnl,Duztas:2019mxr,Song:2018nqg,Duztas:2016xfg,Duztas:2021kuj,Shaymatov:2020wtj,Jiang:2020mws,Gwak:2021tcl}. However, a newer version of checking WCCC by Sorce and Wald \cite{Sorce:2017dst} shows that if the self-force and finite size effects are considered, then no such black hole destruction as illustrated in \cite{Hubeny:1998ga,Jacobson:2009kt,Gao:2012ca,Siahaan:2016zjw} can occur. 

As we know, there exists gravitational theories beyond Einstein-Maxwell. For example, considering an extra dilaton field in addition to the graviton and Maxwell fields, namely the Einstein-Maxwell-dilaton theory. An effective action corresponding to the four dimensional Einstein-Maxwell-dilaton theory can be obtained from the five dimensional vacuum Einstein after performing a Kaluza-Klein compactification. A four dimensional black hole solution in such theory is known as the Kaluza-Klein black hole \cite{Gibbons:1985ac,Larsen:1999pp,Horne:1992zy} which has been studied quite extensively in literature, ranging from those works which may be relevant for some astrophysical purposes \cite{Jai-akson:2017ldo,Amarilla:2013sj,Tsukamoto:2017fxq} to some holography related \cite{Li:2010ch,Azeyanagi:2008kb,Hartman:2008pb,Chen:2010ywa}. 

Interests in these black hole solutions motivate us to test the WCCC associated to them. In the Einstein-Maxwell theory, we investigate the possibility to destroy the near extremal Kerr-Newman black hole in the way that slightly different compared to the one presented in \cite{Gao:2012ca,SaaPRD}. The same way in testing WCCC is also applied to the rotating and charged black hole in Einstein-Maxwell-dilaton theory, in particular the Kaluza-Klein black hole case \cite{Gibbons:1985ac,Horne:1992zy,Larsen:1999pp}. We show that, with the help of numerical results, both black holes can be destroyed by letting them capture test object from far away. Indeed, similar to the presentations in \cite{Gao:2012ca,Siahaan:2015ljs,Siahaan:2016zjw,SaaPRD,Duztas:2019mxr,Duztas:2020xnl}, here we neglect the  self-energy or backreaction factors which recently were used by some authors in proving the WCCC violation proposed in \cite{Hubeny:1998ga,Jacobson:2009kt}. For Kaluza-Klein black hole discussion, we also add an analysis on the naked singularity production by considering the neutral scalar field absorption which leads to an overspinning of the black hole. This consideration has been performed for the Kerr-Newman case in \cite{Duztas:2019ick}.

The organization of this paper is as follows. In section \ref{sec.BH}, we provide a brief review on some features of black hole solutions studied in this paper. In the next section, we obtain the required maximum and minimum energies for the test object which may destroy these black holes. Then in section \ref{sec.Destroy}, we evaluate some numerical examples which illustrate the naked singularity production. To support the result for Kaluza-Klein case in section \ref{sec.Destroy}, overspinning an extremal Kaluza-Klein black hole by a test neutral scalar field is considered in section \ref{sec.scalar}. Finally, we give conclusions. We consider the natural units $c={\hbar} = k_B = G_4 = 1$.

\section{Black hole solutions}\label{sec.BH}

Let us first review the Kaluza-Klein solution of Einstein-Maxwell-dilaton theory. Consider the action for vacuum Einstein in five dimension 
\be \label{action5}
S = \int {{d^5}x\sqrt {\left| {\tilde g} \right|} } \tilde R \,,
\ee 
where $\tilde R$ is the five dimensional Ricci scalar, $\tilde g$ is the determinant of five dimensional spacetime metric $g_{MN}$. Here we used the coordinate $
x^M  = \left\{ {x^\mu  ,y} \right\}$ where the four dimensional spacetime is denoted by coordinate $x^\mu$, and $y$ is the coordinate for fifth dimension. This fifth dimension is compact and has a period of $2\pi R_{5}$. We also assume that the five dimensional spacetime to have the Killing vector $\partial_y$. By using the five dimensional line element
\be \label{metric5}
 ds_5^2  = e^{{{4\Phi } \mathord{\left/
 			{\vphantom {{4\Phi } {\sqrt 3 }}} \right.
 			\kern-\nulldelimiterspace} {\sqrt 3 }}} \left( {dy + 2A_\mu  dx^\mu  } \right)^2  + e^{ - {{2\Phi } \mathord{\left/
 			{\vphantom {{2\Phi } {\sqrt 3 }}} \right.
 			\kern-\nulldelimiterspace} {\sqrt 3 }}} g_{\mu \nu } dx^\mu  dx^\nu \,,
\ee 
the five dimensional action (\ref{action5}) can be written as
\be \label{action4}
S = 2\pi R_5 \int {d^4 x\sqrt {\left| g \right|} \left[ {R - 2\left( {\nabla \Phi } \right)^2  - e^{ - 2\sqrt 3 \Phi } F_{\mu \nu } F^{\mu \nu } } \right]} \,,
\ee 
where $R$ is now the four dimensional Ricci scalar and $g$ is the determinant of four dimensional metric tensor $g_{\mu\nu}$. Above,  $F_{\mu \nu }  = \partial _\mu  A_\nu   - \partial _\nu  A_\mu $ is the field-strength tensor of electromagnetic field.

Taking a variation to the action (\ref{action4}) gives us the equation of motions 
\be\label{EOM1} 
\nabla _\mu  \left( {e^{ - 2\sqrt 3 \Phi } F^{\mu \nu } } \right) = 0\,,
\ee 
\be\label{EOM2} 
\nabla ^2 \Phi  + \frac{{\sqrt 3 }}{2}e^{ - 2\sqrt 3 \Phi } F_{\mu \nu } F^{\mu \nu }  = 0\,,
\ee 
and
\be \label{EOM3}
R_{\mu \nu }  - 2\nabla _\mu  \Phi \nabla _\nu  \Phi  - 2e^{ - 2\sqrt 3 \Phi } \left[ {F_{\mu \lambda } F_\nu ^\lambda   - \frac{1}{4}g_{\mu \nu } F_{\alpha \beta } F^{\alpha \beta } } \right] = 0\,.
\ee 
A rotating and charged black hole solution which solves the equations of motion above is \cite{Horne:1992zy}
\[ 
ds_4^2  = g_{\mu \nu } dx^\mu  dx^\nu   =  - \frac{{1 - Z}}{\beta }dt^2  - \frac{{2aZ\Delta _x }}{{\beta \Delta _v }}dtd\phi  + \beta \Sigma \left( {\frac{{dr^2 }}{{\Delta _r }} + \frac{{dx^2 }}{{\Delta _x }}} \right)
\]
\be \label{metric.KK}
+ \left[ {\beta \left( {\Delta _r  - 2Mr} \right) + \frac{{a^2 \Delta _x Z}}{\beta }} \right]\Delta _x d\phi ^2 \,,
\ee 
where
\be 
\beta = \sqrt {1 + \frac{{v^2 Z}}{{\Delta _v }}} \,,
\ee 
\be 
Z = \frac{2Mr}{\Sigma}\,,
\ee 
$\Delta_r = r^2 - 2Mr + a^2$, $\Sigma = r^2 + a^2 x^2$, $\Delta_x = 1-x^2$, and $\Delta_v = 1-v^2$. Here $v$ is normally referred as the boost velocity parameter \cite{Aliev:2008wv}. The accompanying non gravitational fields are
\be \label{A.kk}
A_\mu  dx^\mu   = \frac{{vZ}}{{2\beta ^2 \Delta _v }}\left( {dt - a\Delta _x \sqrt {\Delta _v } d\phi } \right)
\ee 
for the vector field, and
\be \label{dilaton}
\Phi = -\frac{\sqrt{3}}{2}\beta\,,
\ee 
for the dilaton. 

Note that the black hole spacetime (\ref{metric.KK}) possesses the $\partial_t$ and $\partial_\phi$ Killing symmetries, which is also the case for Kerr-Newman spacetime. This allows us to employ the Komar integral to compute the following conserved quantities, namely the mass
\be \label{Mphys}
{\tilde M} = \frac{{M\left( {2- v^2 } \right)}}{{2\Delta _v }}\,,
\ee 
and the angular momentum
\be \label{Jphys}
{\tilde J} = \frac{Ma}{\sqrt{\Delta_v}}\,.
\ee 
For the conserved electric charge, the standard textbook formula gives us
\be \label{Qphys}
{\tilde Q} = \frac{Mv}{\Delta_v}\,.
\ee 
Here we consider the positive electric charge for the black hole, therefore $0\le v\le 1$. 

As we have mentioned that the spacetime (\ref{metric.KK}) can contain a black hole, there exists horizon with the radius is given by the larger roots of $\Delta_r$, i.e. 
\be 
r_+ = M + \sqrt{M^2 - a^2}\,.
\ee 
It turns out that this is just the outer horizon of Kerr black hole, where $M\ge a$ is the corresponding black hole condition. Accordingly, the extremality is achieved at $a=M$ which yields $r_{\rm ext.} = M$ for a Kaluza-Klein black hole. However, note that mass $M$ and rotational parameter $a$ above are not the proper conserved quantities that associate to a Kaluza-Klein black hole, due to the results in (\ref{Mphys}), (\ref{Jphys}), and (\ref{Qphys}). Using the inverse of these relations for a positive total charge $\tilde{Q}$, namely
\be 
v = \frac{{2\tilde Q}}{{\tilde M + \sqrt {\tilde M^2  + 2\tilde Q^2 } }}\,,
\ee 
\be 
M = \frac{{\tilde M}}{2}\left( {3 - \sqrt {1 + \frac{{2\tilde Q^2 }}{{\tilde M^2 }}} } \right)
\,,
\ee 
and
\be 
a = \frac{{\sqrt 2 \tilde J}}{{\tilde M\sqrt {1 - \frac{{\tilde Q^2 }}{{\tilde M^2 }} + \left( {1 + \frac{{2\tilde Q^2 }}{{\tilde M^2 }}} \right)^{1/2} } }}
\,,
\ee 
the horizon radius in extremal case can be rewritten as
\be
r_{\rm{ext.}} = \frac{{3\tilde M - \sqrt {\tilde M^2  + 2\tilde Q^2 } }}{2}\,,
\ee 
accordingly.

Now let us turn to the Einstein-Maxwell theory whose action can be obtained by setting $\Phi\to 0$ in the Einstein-Maxwell-dilaton action (\ref{action4}) above. The corresponding equations of motion then read
\be \label{eq.source-free}
{\nabla _\mu }{F^{\mu \nu }} = 0\,,
\ee 
and
\be \label{eq.Einstein-Maxwell}
{R_{\mu \nu }} = 2{F_{\mu \alpha }}F_\nu ^\alpha  - \frac{1}{2}{g_{\mu \nu }}{F_{\alpha \beta }}{F^{\alpha \beta }}\,.
\ee 
The rotating and charged black hole spacetime satisfying these equations of motion is known as the Kerr-Newman black hole solution whose line element can be written as
\be \label{metric.KN}
d{s^2} =  - \frac{{{\Delta _{KN}}}}{\Sigma }{\left( {dt - a{\Delta _x}} \right)^2} + \Sigma \left( {\frac{{d{r^2}}}{{{\Delta _{KN}}}} + \frac{{d{x^2}}}{{{\Delta _x}}}} \right) + \frac{{{\Delta _x}}}{\Sigma }{\left( {adt - \left( {{r^2} + {a^2}} \right)d\phi } \right)^2}\,,
\ee 
where
\be {\Delta _{KN}} = {r^2} - 2Mr - {a^2} - {Q^2}\,.
\ee
Note that the functions $\Sigma$ and $\Delta_x$ appearing in the Kerr-Newman metric above are just the same to those in Kaluza-Klein solution (\ref{metric.KK}). On the other hand, the accompanying vector solution in solving the Einstein-Maxwell equations (\ref{eq.source-free}) and (\ref{eq.Einstein-Maxwell}) reads
\be \label{A.kn}
{A_\mu }d{x^\mu } =  - \frac{{{\tilde Q}r}}{\Sigma }\left( {dt - a{\Delta _x}d\phi } \right)\,.
\ee 
Again, as a black hole solution, the Kerr-Newman singularity is covered by an event horizon located at $r_+ = M +\sqrt{M^2-a^2-Q^2}$. The extremal state for Kerr-Newman black hole is denoted by $M^2=a^2+Q^2$, and in this condition we have $r_+ =M$. Moreover, as we have mentioned that the Kerr-Newman metric enjoy the $\partial_t$ and $\partial_\phi$ Killing symmetries, the black hole mass $M$ and angular momentum $J=Ma$ can be verified by using the Komar integral associated with each of these Killing symmetries. Furthermore, the surface integral to obtain total electric charge in the spacetime (\ref{metric.KN}) gives us $Q$ as the charge of black hole\footnote{Here we use the tilde notation for Kerr-Newman black hole charge to make it distinguished to the charge of Kaluza-Klein black hole above.}. 
 
\section{Energies of test particle}\label{sec.Energy}

In this section, we would like to obtain the upper and lower limits for the energy of a test particle that may destroy the black holes under consideration. We imagine that a test particle released from far away and moving towards the black hole. Definitely, the properties of black hole will change after capturing this test object which is parameterized by some physical properties such as energy $E$, charge $q$, and angular momentum $L$. Indeed, since both of the Kerr-Newman and Kaluza-Klein black hole spacetimes possess the $\partial_t$ and $\partial_\phi$ Killing symmetries, $E$ and $L$ are normally known as the two constants of motion in the spacetime. If the change of black hole parameter suggest the production of naked singularity after capturing the test particle, then we can infer that the black hole is already destroyed by the test particle.

As a start, let us first discuss the geodesics of a charged test particle in a general rotating charged black hole stationary spacetime with axial symmetry. This applies to the cases we consider in this paper, namely the Kerr-Newman and Kaluza-Klein black hole spacetimes. In a general curved background, the motion of a massive charged test particle is dictated by the geodesic equation
\be\label{eq.motion.incurved}
{\ddot x}^\mu + \Gamma _{\alpha \beta }^\mu {\dot x}^\alpha {\dot x}^\beta   = \frac{q}{\mu}F^{\mu \nu } {\dot x}_\nu\,,
\ee 
where $\mu$ and $q$ are the mass and electric charge of the particle, respectively. In this geodesic equation, the ``dot'' stands for the derivative with respect to some affine parameter $s$, ${\dot ()} = \tfrac{d()}{ds}$. As we know, it can be shown that the geodesic equation above is associated with the Lagrangian
\be\label{eq.Lang}
{\cal L} = \frac{1}{2}\mu g_{\alpha \beta } {\dot x}^\alpha {\dot x}^\beta + qA_\mu  {\dot x}^\mu\,.
\ee 
From this Lagrangian, the energy $E$ and angular momentum $L$ as the constants of motion for the particle in a stationary background can be obtained as
\be
E =  - \frac{{\partial {\cal L}}}{\partial {\dot t}} =  - \mu\left( {g_{tt} \dot t + g_{t\phi } \dot \phi } \right) - qA_t\, ,
\ee 
and
\be 
L = \frac{{\partial {\cal L}}}{\partial {\dot \phi}} = \mu\left( {g_{t\phi } \dot t + g_{\phi \phi } \dot \phi } \right) + qA_\phi\,.
\ee 

Furthermore, since ${\dot x}_\mu{\dot x}^\mu = -1$ for the timelike object, one can obtain a relation between $E$ and $L$ from the last two equations above that reads
\be \label{EtoL}
E = \frac{{g_{t\phi } }}{{g_{\phi \phi } }}\left( {qA_\phi   - L} \right) - qA_t  + {\sqrt {{\left( {\frac{{g_{t\phi }^2  - g_{\phi \phi } g_{tt} }}{{g_{\phi\phi }^2 }}} \right)\left( {\left( {L - qA_\phi  } \right)^2  + \mu^2 g_{\phi \phi } \left( {1 + g_{rr} \dot r ^2 + g_{\theta \theta } \dot \theta ^2} \right)} \right)}}}\,.
\ee
Note that, in getting eq. (\ref{EtoL}) we have considered the solution that implies ${\dot t} > 0$ only. Subsequently, inserting the corresponding components of $g_{\mu\nu}$ into (\ref{EtoL}) followed by the $r=r_+$ evaluation, the minimum energy for a test particle to reach the event horizon can be obtained. In Kaluza-Klein spacetime, this minimum energy reads\footnote{Note that the mass $m$ appearing in this minimum energy is the one from Kerr seed solution, as one can view Kaluza-Klein black hole spacetime is a product of some transformation with boost velicity parameter $v$ with Kerr metric as the seed \cite{Aliev:2008wv}. The proper conserved mass for Kaluza-Klein black hole is given in (\ref{Mphys}).}
\be\label{Emin-notEx}
E_{\min }  = \frac{{aL\sqrt {\Delta _v }  - vq M r_ +  }}{{2 M r_ +  }}\,,
\ee 
while for the Kerr-Newman case is \cite{Gao:2012ca}
\be 
{E_{\min }} = \frac{{aL + q  Q{r_ + }}}{{{a^2} + r_ + ^2}}\,.
\ee 
Therefore, for a test particle to destroy the black hole, its energy cannot be lower than the minimum energy above for each particular case.

On the other hand, there exist an upper bound for a test particle energy which could destroy a black hole. In other words, if this energy is bigger than a maximum value that is allowed, the black hole destruction can never occur. The maximum energy for the test particle itself comes from an inequality that violates the black hole condition. For the Kaluza-Klein case, the black hole condition reads
\be \label{inq.KK}
\frac{\tilde M}{2} \left( {3 - \sqrt {1 + \frac{{2{\tilde Q^2}}}{{{\tilde M^2}}}} } \right) \ge \sqrt 2 \frac{\tilde J}{\tilde M} {\left( {1 - \frac{{{\tilde Q^2}}}{{{\tilde M^2}}} + \sqrt {1 + \frac{{2{\tilde Q^2}}}{{{\tilde M^2}}}} } \right)^{ - \frac{1}{2}}}\,.\ee 
At this point, for a future benefit, we can define
\be\label{del.kk} 
{\delta _{KK}} \equiv M - a = \frac{\tilde M}{2}\left( {3 - \sqrt {1 + \frac{{2{\tilde Q^2}}}{{{\tilde M^2}}}} } \right) - \sqrt 2 \frac{\tilde J}{\tilde M}{\left( {1 - \frac{{{\tilde Q^2}}}{{{\tilde M^2}}} + \sqrt {1 + \frac{{2{\tilde Q^2}}}{{{\tilde M^2}}}} } \right)^{ - \frac{1}{2}}}\,,\ee
as the near to extremal parameter in Kaluza-Klein black hole case. Obviously, the extremal condition is represented by $\delta_{KK}=0$, and the black hole destruction is denoted by $\delta_{KK} < 0$. Moreover, it will be useful for us later to show the angular momentum of Kaluza-Klein black hole at extremal state, which can be written as
\be 
{\tilde J} = \frac{{\sqrt 2 \left( {3\tilde M - \sqrt {\tilde M^2  + 2\tilde Q^2 } } \right)\left( {\tilde M\sqrt {\tilde M^2  + 2\tilde Q^2 }  + \tilde M^2  - \tilde Q^2 } \right)}}{4}\,.
\label{angular}
\ee 

Obviously, as the black hole captures a test particle with energy $E$, angular momentum $L$, and electric charge $q$, the black hole physical parameters change as ${\tilde M}\to {\tilde M}+E$, ${\tilde J} \to {\tilde J} +L$, and ${\tilde Q} \to {\tilde Q} +q$. Consequently the corresponding black hole condition changes as well, and the related inequality for a naked singularity production reads 
\be \label{inq.KK-2}
{({\tilde M}+E)^2}\left( {3 - \sqrt {1 + \frac{{2{({\tilde Q}+q)^2}}}{{{({\tilde M}+E)^2}}}} } \right) <  2\sqrt 2 ({\tilde J}+L){\left( {1 - \frac{{{({\tilde Q}+q)^2}}}{{{({\tilde M}+E)^2}}} + \sqrt {1 + \frac{{2{({\tilde Q}+q)^2}}}{{{({\tilde M}+E)^2}}}} } \right)^{ - \frac{1}{2}}}\,.\ee
However, solving Eq.~(\ref{inq.KK-2}) to obtain the exact form of $E_{\rm max}$ is troublesome. Thence we omit to do so and rely on some numerical results in the next section to tell whether the Kaluza-Klein black hole can be destroyed or not.

Now let us turn to the Kerr-Newman case where the maximum energy for the test particle is given by the inequality
\be {\left( { M + E} \right)^2} < {\left( { Q + q} \right)^2} + \frac{{{{\left( { J + L} \right)}^2}}}{{{{\left( { M + E} \right)}^2}}}\,,\ee 
i.e.
\be \label{E-max-KN}
{E_{\max }} = \frac{1}{2}{\left( {2{{\left( { Q + q} \right)}^2} + 2\sqrt {{{\left( { Q + q} \right)}^4} + 4{{\left( { J + L} \right)}^2}} } \right)^{\frac{1}{2}}} -  M\,.
\ee 
The corresponding near to extremal parameter for Kerr-Newman black hole is
\be \label{del.KN}
\delta _{KN}^2 = {{ M}^2} - {{ Q}^2} - \frac{{{{ J}^2}}}{{{{ M}^2}}}\,,\ee
and the extremal state is denoted by $\delta _{KN}=0$. Accordingly, solving eq. (\ref{del.KN}) for $J$ and insert the result into eq. (\ref{E-max-KN}) gives us
\begin{equation}
E_{\max} = \frac{(q+{ Q})^2+\sqrt{(q+{ Q})^4 + 4 \left(L+{ M}\sqrt{{ M}^2-{ Q}^2-\delta_{KN}^2}\right)^2}}{\sqrt{2}}
\label{E-max-KN-2} -{ M}\,.
\end{equation}
In the next section we will examine some numerical examples for these $E_{\rm max}$ and $E_{\rm min}$ above to seek a possibility for the black holes to be destroyed. The existence of $E_{\rm max} > E_{\rm min}$ for a range of $\delta$ suggests that the black hole can be destroyed by a test particle.

\section{Destroying the black holes}\label{sec.Destroy}

\begin{figure}
	\centering
	\includegraphics[scale=0.5]{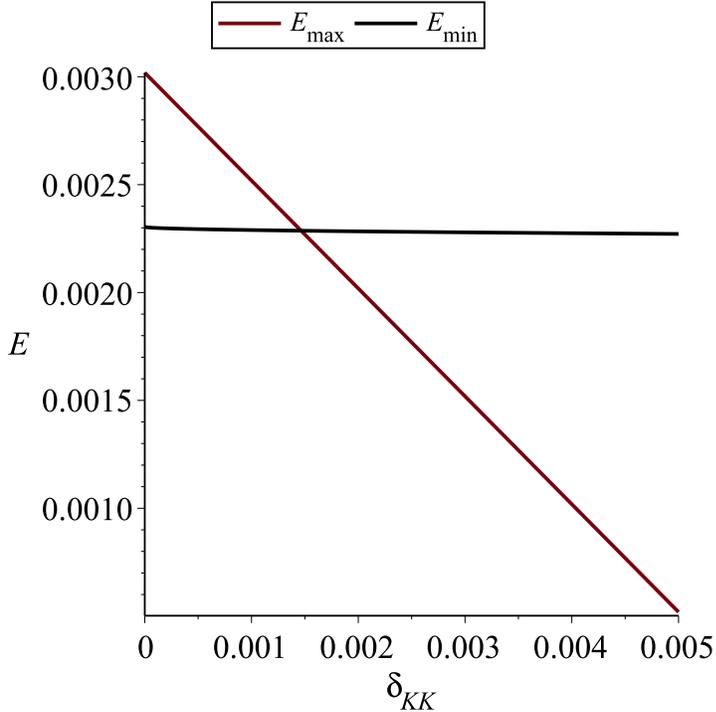}
	\caption{Plot of $E_{\rm max}$ and $E_{\rm min}$ for a KK black hole with a neutral test particle.}\label{fig.Ekk1}
\end{figure}

\begin{figure}
	\centering
	\includegraphics[scale=0.5]{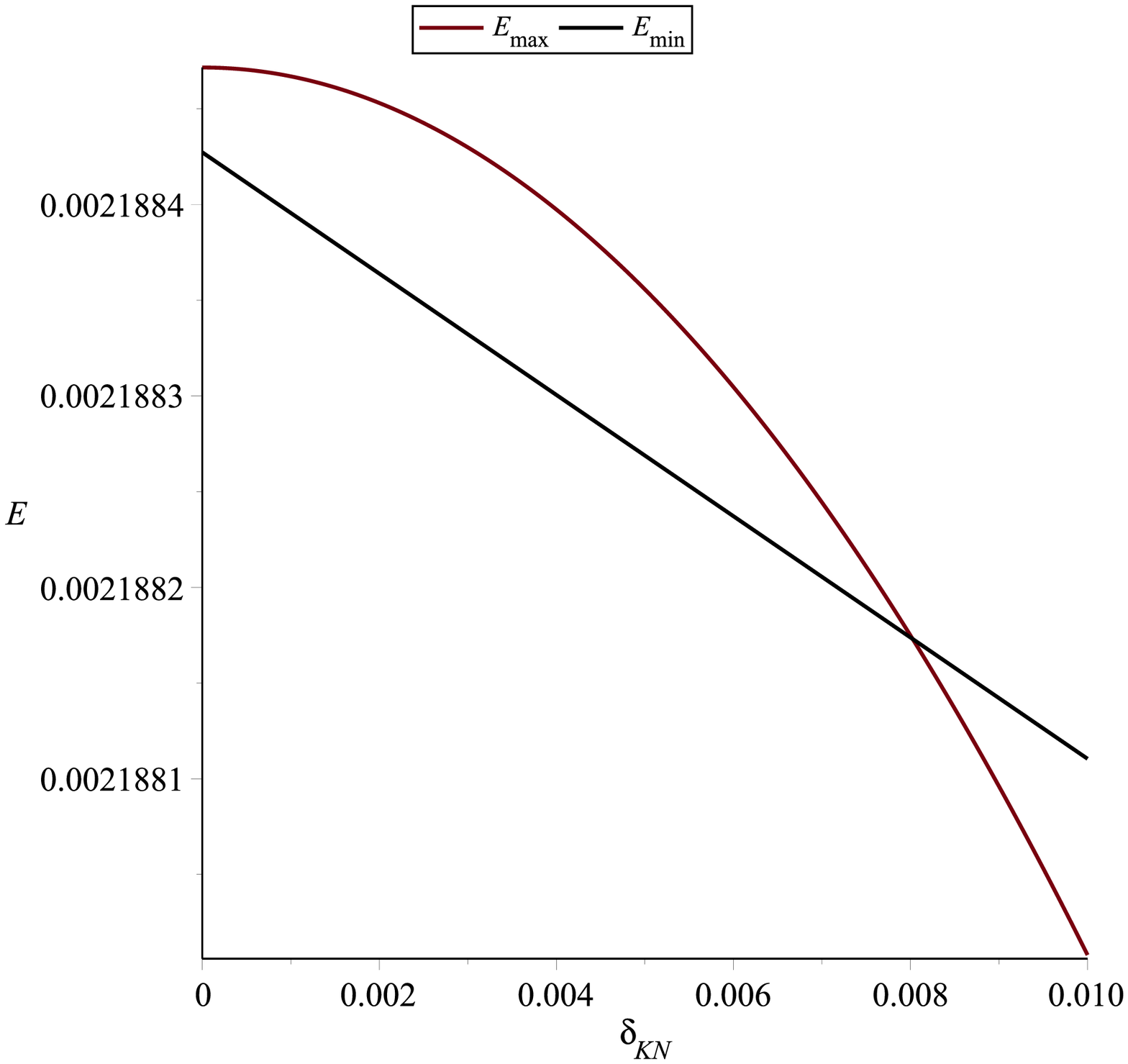}
	\caption{Plot of $E_{\rm max}$ and $E_{\rm min}$ for a KN black hole with a neutral test particle.}\label{fig.Ekn}
\end{figure}

Here we provide some numerical evaluations of eqs. (\ref{inq.KK-2}) and (\ref{E-max-KN-2}) by using some setups which allow the production of a naked singularity from a near extremal black holes. In getting the plots for maximum and minimum energies, we can use the numerical values for $M={\tilde M}=100$, $Q={\tilde Q}=96$, $L=0.5$, and $q=10^{-3}$. The results are given in figs. \ref{fig.Ekk1} and \ref{fig.Ekn} where in both plots we can observe that there is a range of small $\delta$ where $E_{\rm max} > E_{\rm min}$. From this fact we can infer that the near extremal Kaluza-Klein and Kerr-Newman black holes can be destroyed by using a test particle.  

\begin{figure}
	\centering
	\includegraphics[scale=0.5]{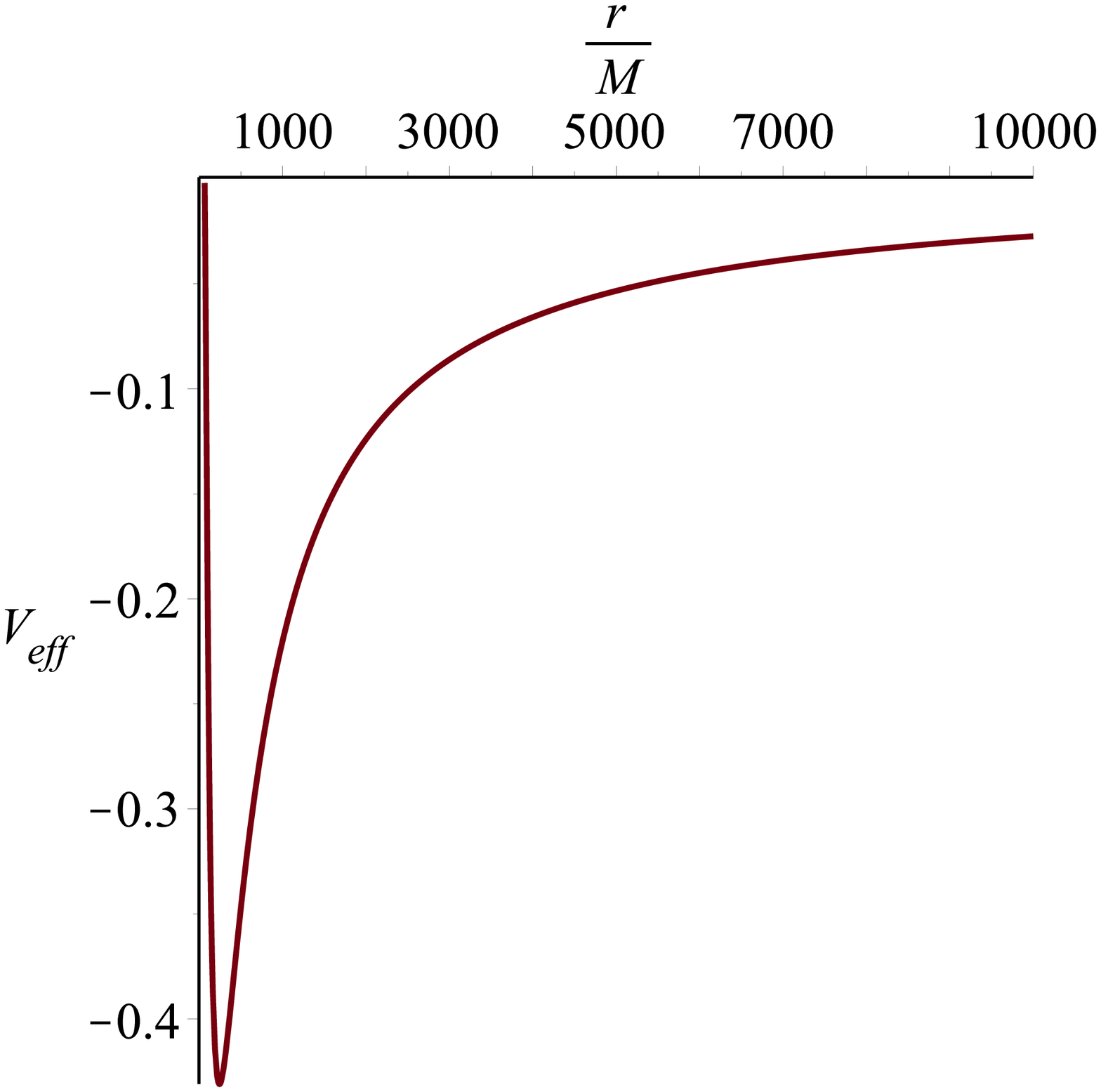}
	\caption{Effective potential for the test particle discussed in Fig.~\ref{fig.Ekk1}.}\label{fig.Vkk}
\end{figure}

\begin{figure}
	\centering
	\includegraphics[scale=0.5]{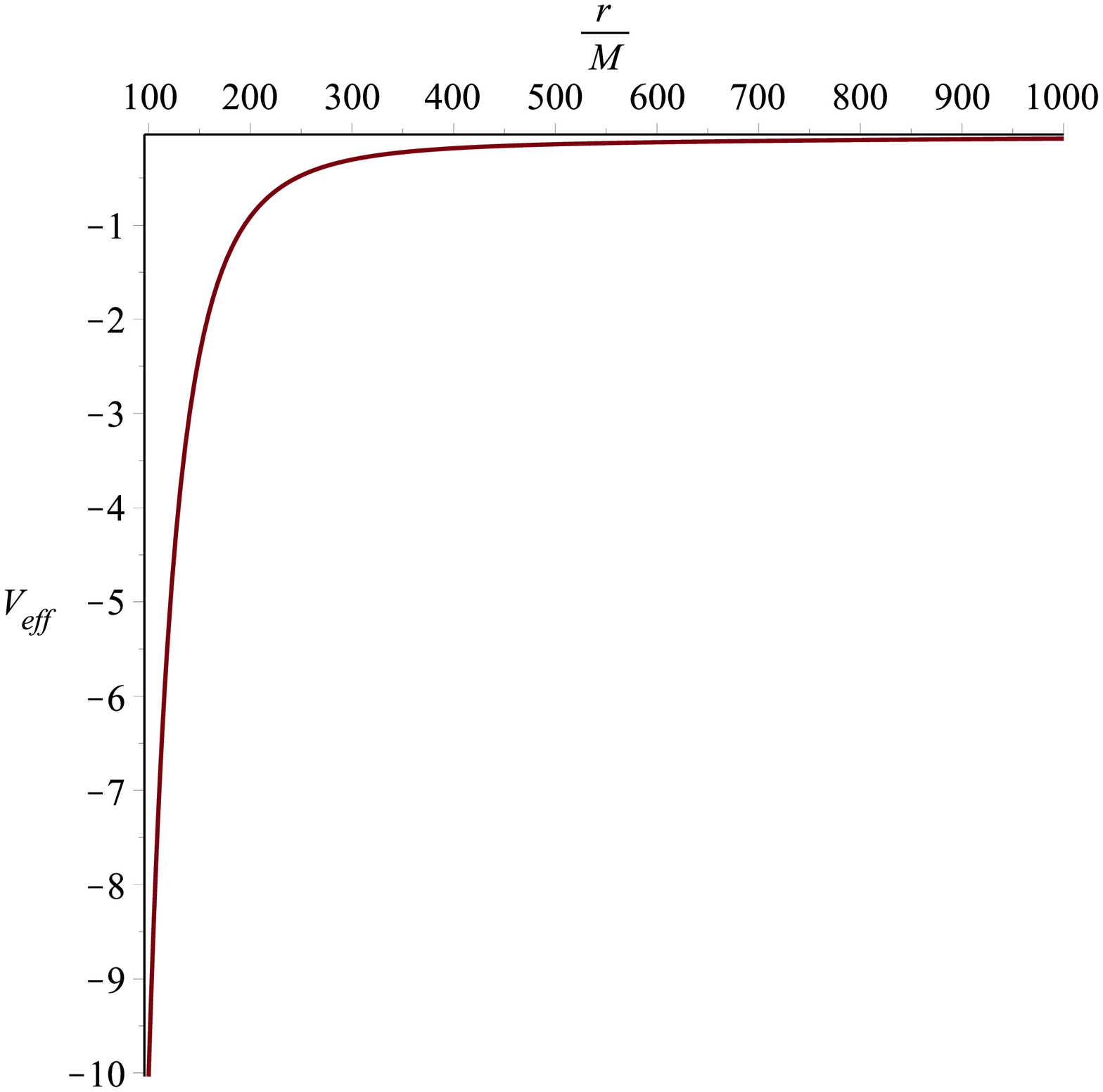}
	\caption{Effective potential for the test particle discussed in fig. \ref{fig.Ekn}.}\label{fig.Vkn}
\end{figure}

Now we need to confirm that the test particle that fits the description in figs. \ref{fig.Ekk1} and \ref{fig.Ekn} can really reach the horizon. For this purpose, we can use the effective potential $V_{eff} = - {\dot r}^2$ method, namely if we can show that $V_{eff} <0$ along the path then we can assure that the particle can really fall into the black hole from far distance. To proceed, we can pick an appropriate case based on the result in figs. \ref{fig.Ekk1} and \ref{fig.Ekn}, for example $\delta_{KK} = \delta_{KN} = 5\times 10^{-4}$. For the Kaluza-Klein black hole case, this implies $E_{\min} = 0.002293799446$ and $E_{max} = 0.00276891218$. Therefore we are allowed to consider a test particle with $E = 0.0024$ released from far away whose effective potential is given in fig. \ref{fig.Vkk}. On the other hand, the associated energies for the Kerr-Newman case are $E_{\min} = 0.002188411458$ and $E_{max} = 0.0021885$. Accordingly we can consider the test particle with $E=0.218845$, which leads to fig. \ref{fig.Vkn}. Both effective potential plots \ref{fig.Vkk} and \ref{fig.Vkn} suggest that the test particle under considerations can really touch the horizon after released from far away.

\begin{figure}
	\centering
	\includegraphics[scale=0.5]{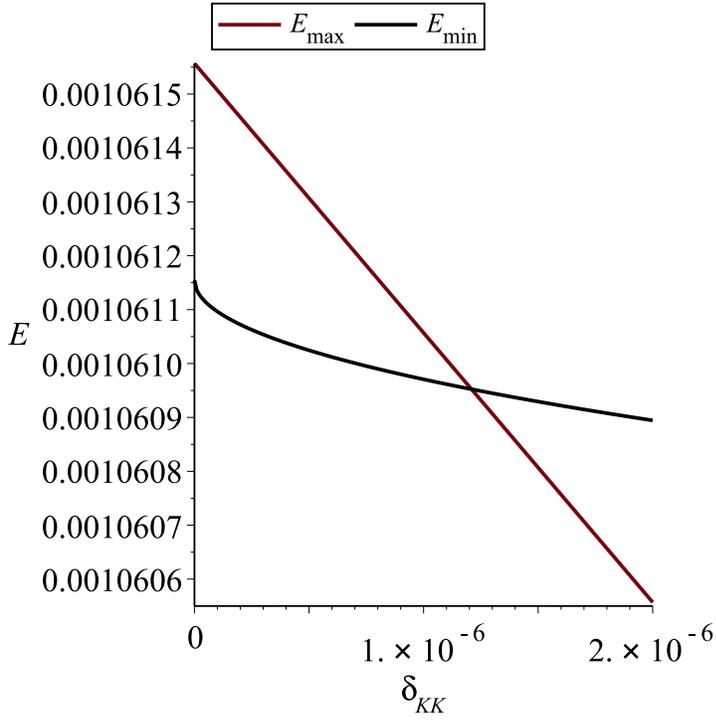}
	\caption{Plot of $E_{\rm max}$ and $E_{\rm min}$ for a KK black hole. }\label{fig.Ekkneu}
\end{figure}

\begin{figure}
	\centering
	\includegraphics[scale=0.5]{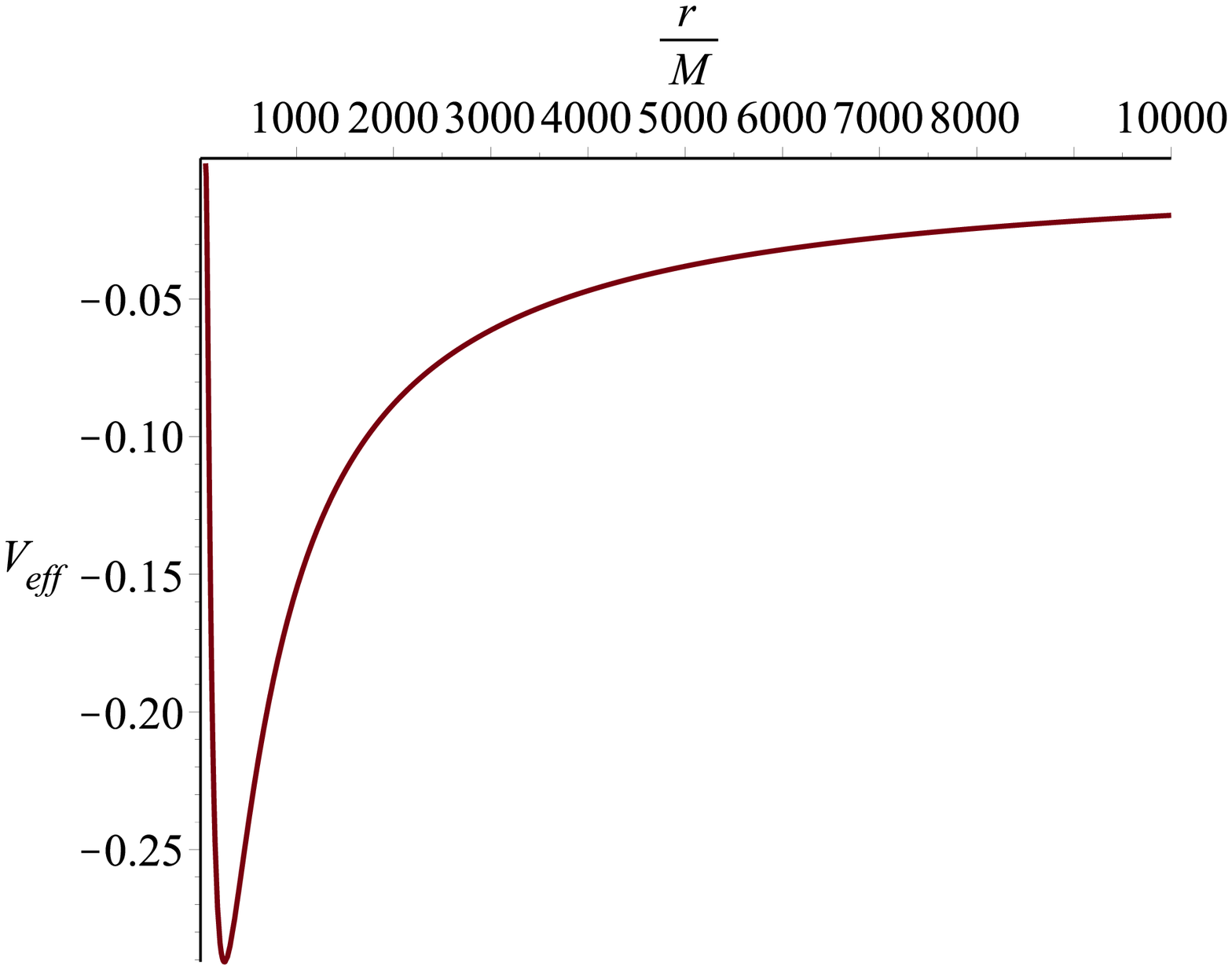}
	\caption{Effective potential for the test particle discussed in Fig.~\ref{fig.Ekk1}.}\label{fig.Vkkneu}
\end{figure}

\begin{figure}
	\centering
	\includegraphics[scale=0.5]{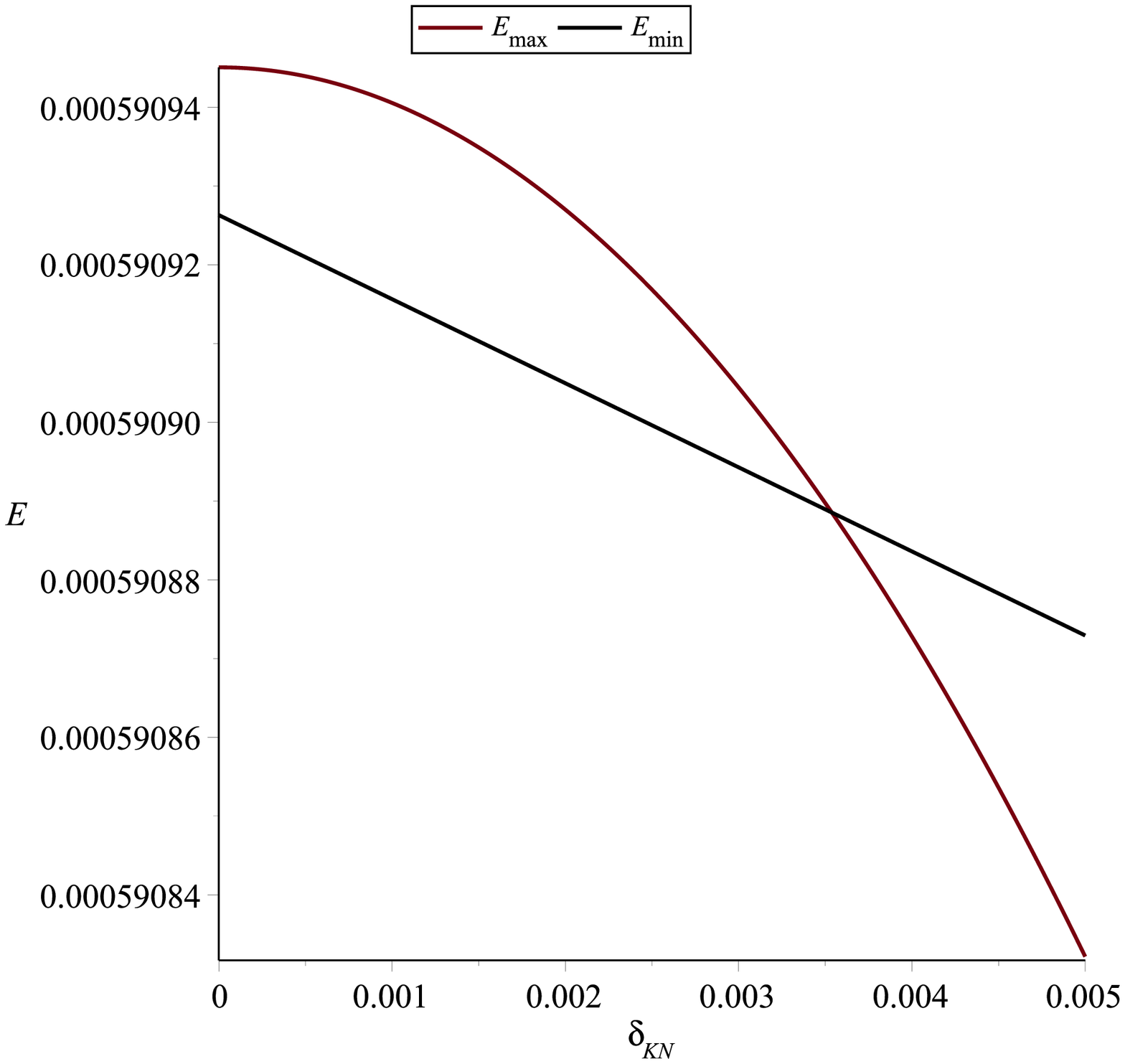}
	\caption{Plot of $E_{\rm max}$ and $E_{\rm min}$ for a KN black hole. }\label{fig.Eknneu}
\end{figure}

\begin{figure}
	\centering
	\includegraphics[scale=0.5]{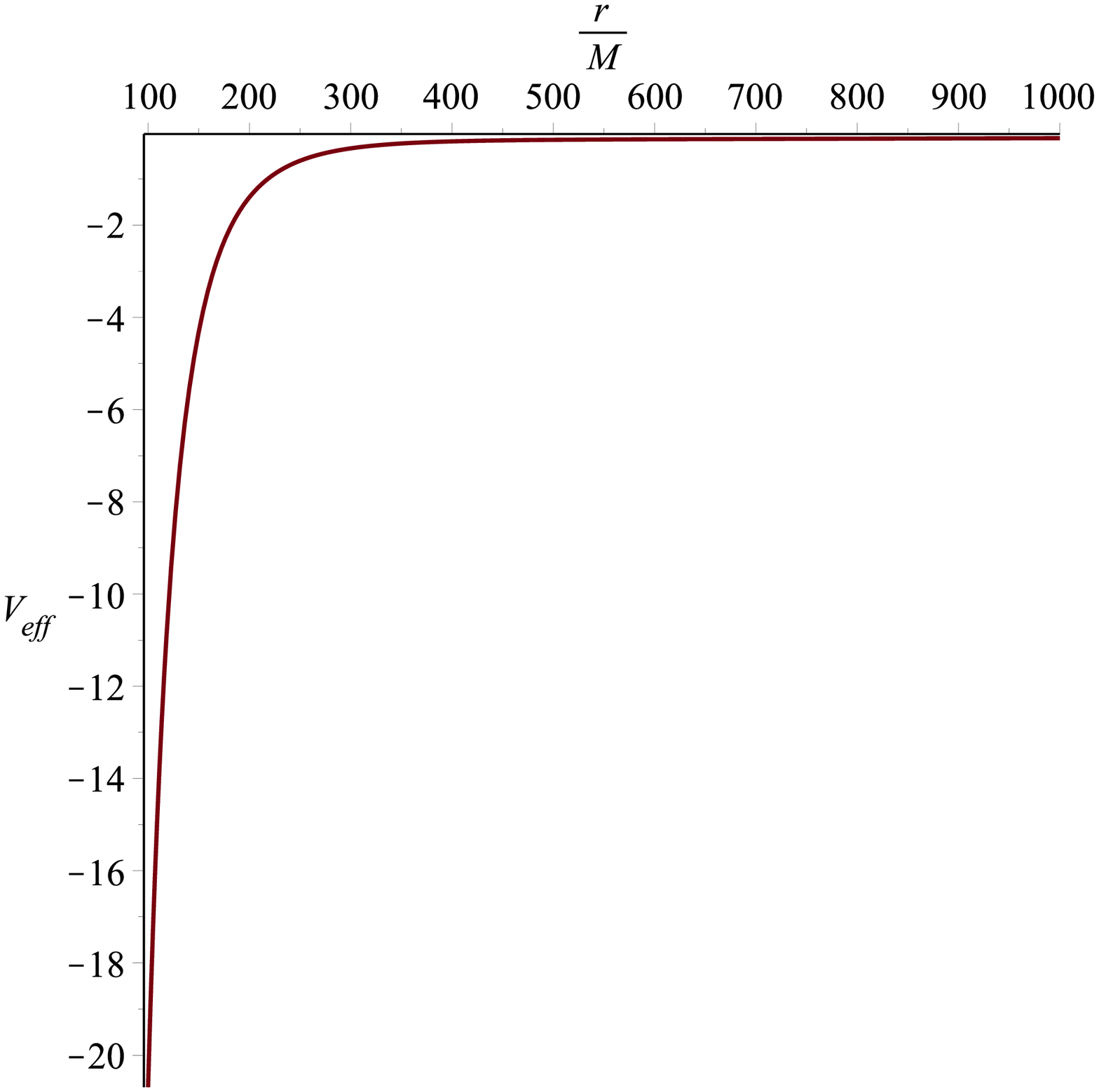}
	\caption{Effective potential for the test particle discussed in fig. \ref{fig.Ekn}.}\label{fig.Vknneu}
\end{figure}

Furthermore, we can also consider the destroying of black holes by a neutral test particle. We can consider the same setup for test object in both backgrounds of Kaluza-Klein and Kerr-Newman black holes, namely $L=0.2$ and $q=0$, and set for the black hole parameters as $M={\tilde M}=100$ and $Q={\tilde Q}=94.5$. We find using this numerical values, both near extremal black holes still can be destroyed by the test particle, as indicated in figs \ref{fig.Ekkneu} and \ref{fig.Eknneu}, where the corresponding effective potentials are given in figs. \ref{fig.Vkkneu} and \ref{fig.Vknneu}.

\section{Scalar field perturbation}\label{sec.scalar}

To support the results in previous section where we have showed that both black holes under consideration can be destroyed, here we provide an investigation on extremal Kaluza-Klein black hole destruction by using a neutral test scalar field. Similar study for Kerr-Newman has been reported in \cite{Duztas:2019ick}, where D\"uzta\c{s} has shown that neutral test scalar field can destroy an extremal Kerr-Newman black hole. The argument is based on the fact that superradiance can occur for a range of frequency $0 < \omega < m \Omega_H^{\rm ext.}$ in the background of extremal rotating black holes, including Kerr-Newman and Kaluza-Klein. Here, $m$ is an integer related to the spheroidal harmonics of the scalar field, and $\Omega_H^{\rm ext.}$ is the angular velocity at horizon. If the frequency is outside of this range, i.e. $\omega > m\Omega_H^{\rm ext.}$, then the field is absorbed by the black hole leading to some changes to the black holes properties. Moreover, one can also show that there exist an upper bound for the frequency of scalar field, namely $\omega_{\rm{max-ext.}}$, to allow the overspinning of an extremal black hole leading to a naked singularity production. Therefore, if we can show that $m \Omega_H^{\rm ext.} < \omega_{\rm{max-ext.}}$ for some particular considerations, then a scalar field with the frequency $m \Omega_H^{\rm ext.} < \omega < \omega_{\rm{max-ext.}}$ can destroy the black hole.

The superradiance effect around Kaluza-Klein black hole has been investigated in \cite{Koga:1994np}. It was shown that superradiance can occur for Kaluza-Klein black hole for the incoming neutral test scalar field with frequency $0 < \omega < \omega_{\rm{sl-ext.}}$, where $\omega_{\rm{sl-ext.}} = m\Omega_H^{\rm ext.}$ and the extremal angular velocity at horizon is given by \cite{Koga:1994np}
\be \label{w-sl-ext}
\Omega_H^{\rm ext.} =  {\frac {\sqrt{2} \left( 1-{k}^{2}+\alpha \right)^{3/2} \left( 3-\alpha \right) }{2{\tilde M}\left(1+\alpha\right)\left(2+2\alpha-10{k}^{2}+4\alpha{k}^{2}-{k}^{4} \right) }}\,.
\ee 
Above, we have expressed $\Omega_H^{\rm ext.}$ in the conserved mass $\tilde M$, rotational parameter $a$, and charge $\tilde Q$. Moreover, we have used the constant $k$ as the ratio of total charge $\tilde Q$ to the black hole mass $\tilde M$, i.e. ${\tilde Q} = k {\tilde M}$, and $\alpha = \sqrt{1 + 2k^2}$ for simplicity. To be explicit, a neutral test scalar field with frequency $ 0 < \omega < \omega_{\rm{sl-ext.}}$ in the background of extremal Kaluza-Klein black hole can gain some energy from the black hole, and those with $\omega > \omega_{\rm{sl-ext.}}$ will be absorbed by the black hole. 

Now we can obtain an upper bound for the frequency of scalar field which allow the production of naked singularity from an extremal Kaluza-Klein black hole using the inequality\footnote{Note that this inequality is just eq. (\ref{inq.KK-2}) with $q=0$.}
\be \label{inq.field}
{({\tilde M}+\delta {\tilde M})^2}\left( {3 - \sqrt {1 + \frac{{2{{\tilde Q}^2}}}{{{({\tilde M}+\delta {\tilde M})^2}}}} } \right) < \frac{2\sqrt 2 ({\tilde J}+\delta {\tilde J})}{\sqrt{{1 - \frac{{{{\tilde Q}^2}}}{{{({\tilde M}+\delta {\tilde M})^2}}} + \sqrt {1 + \frac{{2{{\tilde Q}^2}}}{{{({\tilde M}+\delta {\tilde M})^2}}}} }}}\,.\ee 
Note that we have set $\delta {\tilde Q} = 0$ in the inequality above since the incoming scalar field is neutral. Following \cite{Duztas:2019ick}, we can set $\delta {\tilde J} = m \delta {\tilde M} \omega^{-1}$ and $\delta {\tilde M} = \epsilon {\tilde M}$ for $\epsilon \ll 1$ as the angular momentum and energy of incoming test scalar field, respectively. Solving the last inequality for $\omega$ gives us
\be \label{w-max-ext}
\omega  < \omega _{\rm{max  - ext.}}  = \frac{{2\sqrt 2 m\epsilon }}{{{\tilde M}\left[ {\left( {\alpha  - 3} \right){\cal Y} + \left( {3\epsilon  - \delta  + 3} \right){\cal Z}} \right]}}
\ee 
where $\delta = \sqrt{1+2\epsilon + \epsilon^2 + 2k^2}$, ${\cal Y} = \sqrt{1-k^2+\beta}$, ${\cal Z} = \sqrt{\epsilon^2 + 2 \epsilon + \delta\epsilon +1+\delta-k^2}$, and $\beta = \sqrt{1+2k}$. 

\begin{figure}
	\centering
	\includegraphics[scale=0.5]{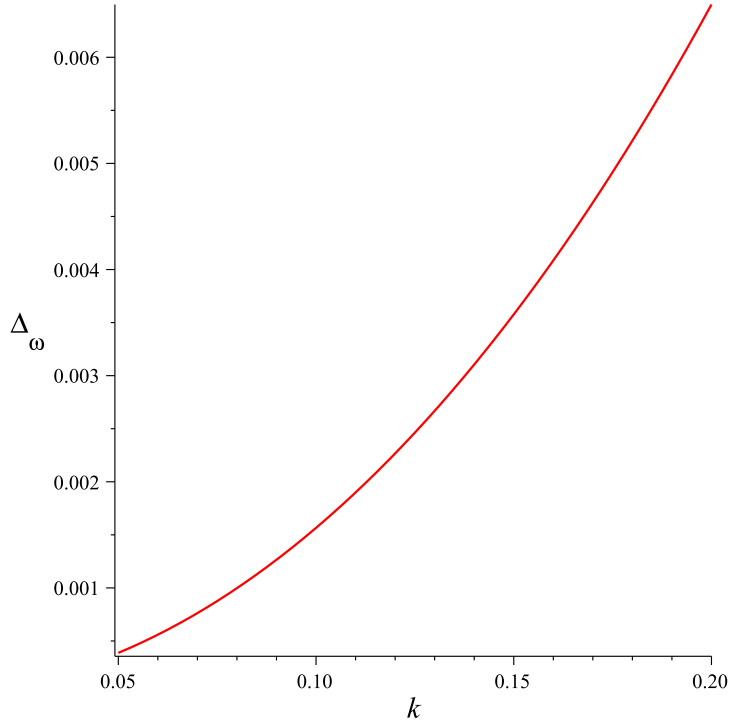}
	\caption{Plot of $\Delta_{\omega}$ vs. $k$ for $\epsilon = 10^{-9}$. }\label{fig.Dw}
\end{figure}

Obviously, the incorporating expressions for $\omega _{\rm{max  - ext.}}$ and $\omega _{\rm{sl  - ext.}}$ in eqs. (\ref{w-max-ext}) and (\ref{w-sl-ext}) above are considerably more complicated compared to that of Kerr-Newman case reported in \cite{Duztas:2019ick}. This hinders us to obtain some exact bound for $k$ allowing $\omega _{\rm{max  - ext.}} > \omega _{\rm{sl  - ext.}}$ in justifying the naked singularity production. Nevertheless, we can define a quantity
\be 
\Delta_{\omega} = \frac{{\tilde M}\left(\omega _{\rm{max  - ext.}} - \omega _{\rm{sl  - ext.}}\right)}{\sqrt{2}{ m}}\,,
\ee 
which measures the gap between $\omega _{\rm{max  - ext.}} $ and $ \omega _{\rm{sl  - ext.}}$, and check it numerically for some appropriate numerical examples. The result of $\Delta_{\omega} > 0$ tells us that the absorption of a test scalar field leading to naked singularity production is possible for an extremal Kaluza-Klein black hole. A numerical example is given in fig. \ref{fig.Dw} where we find there exist $\Delta_{\omega} >0$ for a range of $k$. This result support our previous claim that a Kaluza-Klein black hole can be destroyed.

\section{Conclusions}\label{sec.Conclusion}

In this paper, we have studied the naked singularities production from Kerr-Newman and Kaluza-Klein black holes. Both black holes are charged and rotating, hence the destruction can be done by overspinning or overcharging them. In section \ref{sec.Destroy}, we provide some numerical results for the maximum and minimum of a test body which may lead to horizon destruction. Indeed, further numerical evaluations of the effective potentials for some possible setups leading to black hole destruction are needed, as we have provided in that section. The effective potentials that we obtain confirm that the test body which can destroy the black hole can really fall into the horizon from far away.  

To support the results in section \ref{sec.Destroy}, section \ref{sec.scalar} is devoted to study the overspinning of an extremal Kaluza-Klein black hole by a neutral scalar field. This section is motivated by the work in \cite{Duztas:2019ick} where the author showed how to overspin an extremal Kerr-Newman black hole by using a neutral scalar field. However, due to the complexity of incorporated properties for the Kaluza-Klein black holes, we cannot provide an analytic expression for the ratio between black hole charge $\tilde Q$ to its mass $\tilde M$ which allows the destruction. The results presented in section \ref{sec.scalar} is limited to some numerical examples which suggest that an extremal Kaluza-Klein black hole can be overspun by a scalar field.

Nowadays, there is a growing interest in checking the weak cosmic censorship conjecture by using the method developed by Wald and Sorce \cite{Sorce:2017dst} and it has been applied to some various cases \cite{Li:2020smq,Jiang:2019soz,Jiang:2019vww}. Using this approach, it can be shown that a black hole cannot be overspun or overcharged leading to the violation of weak cosmic censorship conjecture. Performing this method by Wald and Sorce to Kaluza-Klein black hole should be an interesting project and we address it to our future work. 

\section*{Acknowledgement}

This work is supported by Lembaga Penelitian Dan Pengabdian Kepada Masyarakat, Parahyangan Catholic University.

\end{document}